\documentclass[journal]{IEEEtran}
\IEEEoverridecommandlockouts
\usepackage{cite}
\usepackage{amsmath,amssymb,amsfonts}
\usepackage{algorithmic}
\usepackage{graphicx}
\usepackage{textcomp}
\usepackage{xcolor}
\usepackage{comment}
\usepackage[caption=false]{subfig}
\usepackage{url}
\usepackage{soul}
\usepackage[english]{babel}
\usepackage{rotating}
\usepackage{array}
\usepackage[switch]{lineno}
\usepackage{soul}

\def\BibTeX{{\rm B\kern-.05em{\sc i\kern-.025em b}\kern-.08em
    T\kern-.1667em\lower.7ex\hbox{E}\kern-.125emX}}

\title{The Arousal video Game AnnotatIoN (AGAIN) Dataset}

\author{
    \IEEEauthorblockN{David Melhart,
    Antonios Liapis and
    Georgios N. Yannakakis, \emph{IEEE Senior Member}}\\
    \IEEEauthorblockA{\textit{Institute of Digital Games, University of Malta}\\
    Msida, Malta \\
    david.melhart@um.edu.mt, antonios.liapis@um.edu.mt, georgios.yannakakis@um.edu.mt}
}

\begin{document}

\maketitle

\begin{abstract}
How can we model affect in a general fashion, across dissimilar tasks, and to which degree are such general representations of affect even possible? To address such questions and enable research towards \emph{general} affective computing, this paper introduces The Arousal video Game AnnotatIoN (AGAIN) dataset. AGAIN is a large-scale affective corpus that features over 1,100 in-game videos (with corresponding gameplay data) from nine different games, which are annotated for arousal from 124 participants in a first-person continuous fashion. Even though AGAIN is created for the purpose of investigating the generality of affective computing across dissimilar tasks, affect modelling can be studied within each of its 9 specific interactive games. To the best of our knowledge AGAIN is the largest---over 37 hours of annotated video and game logs---and most diverse publicly available affective dataset based on games as interactive affect elicitors.
\end{abstract}


\section{Introduction}

A core challenge of affective computing (AC) is the investigation of \emph{generality} in the ways emotions are elicited and manifested, in the annotation protocols designed, and ultimately in the affect models created. To examine the degree to which general representations of affect are possible and meaningful, AC research requires access to corpora containing affect responses and annotations across dissimilar tasks, participants and annotators. Traditional large-scale AC datasets feature affect annotation of static images, videos, sounds and speech files within a narrow context through which affect is elicited from a particular task. Even when the various tasks under annotation may vary, those are still limited to a very \emph{specific} context---such as viewing a set of social interactions under a theme or playing sessions of the same game.

This paper identifies games as a unique opportunity in AC to observe emergent emotions in a well-defined but highly interactive environment. Interactivity is especially important for the future of AC research as emotions permeate our daily interactions---not just with each other, but with our environment and computers as well. Affective states arising from these interactions impact our behaviour and decision making on a fundamental level \cite{prinz2004gut,damasio1994descartes}. Therefore, modelling emotions that emerge from interactions is becoming paramount to AC research.

\begin{figure}[!tb]
    \centering
    \includegraphics[width=1.0\linewidth]
    {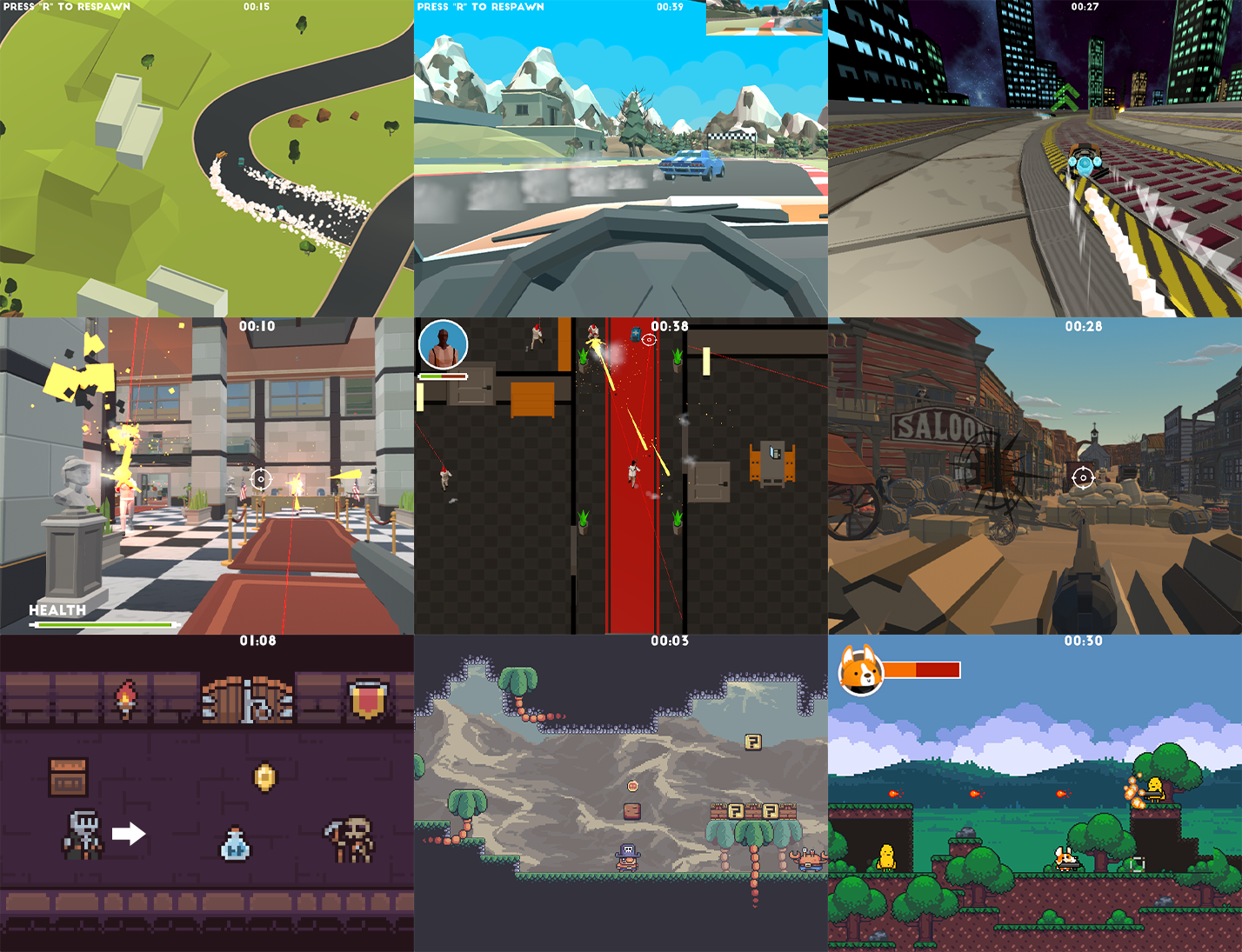}
    \caption{All games featured in the AGAIN dataset currently. The dataset includes 3 racing games (top row), 3 shooter games (middle row), and three platformers (bottom row).}
    \label{fig:all_games}
\end{figure}

Motivated by the lack of corpora for the study of general properties of affect across tasks and participants, in this paper we introduce The Arousal video Game AnnotatIoN (AGAIN) dataset, which contains data from over 120 participants who played and annotated over $1,000$ gameplay sessions. AGAIN is accessible online\footnote{\url{http://again.institutedigitalgames.com/}}
and features data collected from nine games of three dissimilar genres, which were developed specifically for the purposes of the dataset (see Fig. \ref{fig:all_games}). As shown in Table \ref{tab:summary}, along with game telemetry and self-annotated \emph{arousal} labels, the dataset also features a video database of unique gameplay sessions with over 37 hours of in-game footage. The diverse nature of the AGAIN affect elicitors (games) provides a testbed for general affect detection in games \cite{togelius2016general,camilleri2017towards} and broadens the horizons for research on general-purpose AI representations \cite{makantasis2019pixels,mnih2015human} and artificial general intelligence.

\begin{table}[!tb] 
    \caption{Core Properties of the AGAIN Dataset}
    \centering
    \begin{tabular}{|l|c|c|} 
        \hline
        Properties & Raw dataset & Clean dataset\\ \hline\hline
        Number of Participants & 124 & 122\\ \hline
        Number of Gameplay Videos & 1116 & 995\\ \hline
        Number of Game-telemetry Logs & 1116 & 995\\ \hline
        Video database size & 37+ hours & 33+ hours\\ \hline
        Number of Elicitors & \multicolumn{2}{c|}{9 games (3 genres)}\\ \hline
        Gameplay/Video duration & \multicolumn{2}{c|}{2 min}\\ \hline
        Annotation Perspective & \multicolumn{2}{c|}{First-person}\\ \hline
        Annotation Type & \multicolumn{2}{c|}{Continuous unbounded}\\ \hline
        Affective Labels & \multicolumn{2}{c|}{Arousal}\\ \hline 
    \end{tabular}
    \label{tab:summary}
\end{table}

While AC datasets in general rely on collecting peripheral physiological signals in laboratory settings, the AGAIN dataset moves data collection to an online setting. On the one hand, this setup only allows us to collect \emph{behavioural} data in a reliable way. However, since the tools and pipelines employed to collect the dataset emphasise a simple crowdsourced setup, the AGAIN database is much more flexible, extensible and scalable. The design and creation of AGAIN was indeed guided by the following factors: a) \emph{accessibility}, which is achieved through an online crowdsourcing framework; b) \emph{scalability}: AGAIN is utilising the PAGAN online annotation framework \cite{melhart2019pagan} and, hence, one can easily populate the AGAIN database with more participants and annotators; c) \emph{extensibility}: more affect dimensions and categories can be considered and integrated to the existing dataset through the customisable PAGAN annotation tool; d) \emph{generality}: any additional online game or interactive session can be easily integrated to the experimental protocol of AGAIN. While at the time of writing the database hosts 9 games annotated for arousal, AGAIN is designed with all aforementioned factors in mind so that it can host data from more games and user modalities, considering alternative affective labels. 

The AGAIN dataset is unique in a number of ways. First, it is the largest and most diverse publicly available affective dataset based on games as interactive elicitors. Given the breadth of elicitors offered, the dataset can be used for testing specific affect models on one particular task (i.e. a particular game) all the way to general models of affect across tasks (game genres and games in general).  Second, the dataset is annotated with the core affective dimension of arousal, linking dominant annotation practices in affective computing with player modelling and game user research. Finally, it employs a novel annotation framework \cite{lopes2017ranktrace} which captures subjective annotations in a continuous and unbounded manner that can be further processed as labels for regression, classification or ordinal learning affect modelling tasks \cite{yannakakis2018ordinal,yannakakis2017ordinal}.

The remainder of the paper is structured as follows. Section \ref{sec:background} contextualises the dataset within the fields of affective computing and affect modelling in games while Section \ref{sec:affectiveDatasets} offers a systematic review of existing audiovisual datasets. The games used as the affect elicitors of AGAIN are described in Section \ref{sec:games}. Section \ref{sec:annotation} details the AGAIN dataset by describing the protocol followed, the characteristics of the participants, the data types collected, and the annotation framework used. Section \ref{sec:dataset} offers a detailed yet preliminary data analysis of the dataset. Limitations and extensions of AGAIN are discussed in Section \ref{sec:discussion} and the paper concludes with Section \ref{sec:conclusion}.

\section{Background\label{sec:background}}

AGAIN is an accessible dataset offered for research in affective computing at large and player modelling in particular. This background section discusses the importance of arousal within the field of affect representation (Section \ref{sec:pad}) and reviews studies for modelling the affect of game users (i.e. players) in Section \ref{sec:generalPM}.

\subsection{Arousal as Affect Representation}
\label{sec:pad}
While there are different approaches to affect representation including categorical \cite{diehl2007ekman,westbury2015avoid}, dimensional \cite{russell1980circumplex}, and mixed \cite{cambria2012hourglass} frameworks, the AGAIN dataset uses a dimensional representation based on the Pleasure-Arousal-Dominance (PAD) model of affect \cite{mehrabian1980basic} and the Circumplex Model of Emotions \cite{russell1980circumplex}. In contrast to categorical frameworks, which assume a clear division between emotional responses, these models propose a more ambiguous and general representation. Instead of complex emotions, the PAD model focuses on basic affective states represented across three dimensions. \emph{Pleasure} is associated with the valence of the emotion; psychological \emph{arousal} describes the intensity of the emotion; and finally \emph{dominance} describes the agency or level of autonomy during the emotional episode. One can place different emotions within this 3D continuous space without explicitly categorising them, reducing the chance of misrepresenting how a subject feels. This type of evaluation lends itself better for continuous and subjective annotation \cite{yannakakis2018ordinal,yannakakis2017ordinal}. 

While the Circumplex model and the PAD model represent affect across two and three dimensions, respectively, in the AGAIN dataset we focus currently on soliciting annotations based on the dimension of \emph{arousal}. Selecting and investigating arousal first---instead of other affect dimensions---is relevant for games, the core domain of AGAIN. Arousal is present and dominant as an emotional manifestation in game affect interactions and has been associated with challenge \cite{klarkowski2016psychophysiology}, cognitive and affective engagement \cite{abbasi2019empirical}, tension \cite{lopes2017modelling}, fun \cite{clerico2016biometrics}, frustration \cite{melhart2018towards} and flow \cite{seger2012personality}, as well as positive post-game outcomes, such as increased creativity \cite{yeh2015exploring} and working memory \cite{gabana2017effects} performance.
Focusing on one affect dimension reduces the cognitive load of the annotation task \cite{melhart2019pagan}, which in turn increases the reliability of our data; however, it limits the expressive range of affect annotation in the dataset.
Moreover, the focus on arousal assists the research community to build, extend upon and advance research that already has benchmarked the study of arousal in games \cite{lopes2017ranktrace, camilleri2017towards, makantasis2019pixels}.

\subsection{Affect Modelling in Games} \label{sec:generalPM}

Player modelling is the study of video game play both in terms of behavioural and affective patterns \cite{yannakakis2018artificial}. It relies heavily on artificial intelligence methods for building predictive models of player behaviour \cite{bakkes2012player,pfau2019deep}, playtime \cite{mahlmann2010predicting}, churn \cite{perianez2016churn,viljanen2018playtime}, or player experience \cite{makantasis2019pixels,yannakakis2018ordinal,melhart2019your}. It is naturally characterised by dynamic representations and modelling of data, thereby providing even moment-to-moment predictions of a game's elicited experience \cite{melhart2020moment}. A key limitation of player modelling, as with any other data-driven approach, is that it is data hungry. In particular, studies that focus on affective aspects of player experience require ground-truth affect labels which are often costly to collect \cite{wu2016review, d2018affective}.

To address the above challenge, an increasing number of studies focus on approaches that could realise aspects of \emph{general player modelling} \cite{togelius2016general}. General player modelling features methods that are able to predict a player's affective state on unseen games. While early studies such as that of Martinez et al. \cite{martinez2011generic} investigated game-independent features of the playing experience, such as heart rate and skin conductance, later studies put an emphasis on finding general gameplay features either manually \cite{shaker2015towards} or through algorithmic feature mapping \cite{shaker2016transfer}. More recently, Camilleri et al. investigated general gameplay features and generalised metrics of player experience across three dissimilar games \cite{camilleri2017towards}. Their study used high-level features such as \emph{goal-oriented} and \emph{goal-opposed} gameplay events and relative metrics of arousal to moderate success, showing the difficulty of creating general player models. Similarly, Bonometti et al. used high-level general features to characterise the gameplay context (such as activity count and activity diversity) to model engagement across six games published by Square Enix Ltd. \cite{bonometti2020theory}. 

\section{Audiovisual Affective Datasets}
\label{sec:affectiveDatasets}

\begin{sidewaystable*}[ph!]
\centering
\caption{A Survey of Affective Datasets of Audiovisual Content. A table entry is indicated with `N/A' and `UNK' if it is not available and unknown, respectively.}
\label{tab:datsets}
    \resizebox{1\textwidth}{!}{\begin{tabular}{|>{\centering\arraybackslash}m{1.5cm}||>{\centering\arraybackslash}m{1.5cm}|c|>{\centering\arraybackslash}m{1cm}|>{\centering\arraybackslash}m{1cm}||c|>{\centering\arraybackslash}m{3cm}||>{\centering\arraybackslash}m{2cm}|>{\centering\arraybackslash}m{1.7cm}|>{\centering\arraybackslash}m{4cm}|>{\centering\arraybackslash}m{2.2cm}|c|}
        \hline
        & \multicolumn{4}{c||}{\textbf{Elicitation}} & \multicolumn{2}{c||}{\textbf{Participants}} & \multicolumn{5}{c|}{\textbf{Annotation}}\\
        \hline
        \textbf{Database} & \textbf{Interactive} & \textbf{Type} & \textbf{Items} & \textbf{Video} & \textbf{Number} & \textbf{Modalities} & \textbf{Perspective} & \textbf{Type} & \textbf{Labels} & \textbf{Annotators} & \textbf{Tasks} \\
        \hline\hline
        MAHNOB-HCI \cite{lichtenauer2011mahnob} & No & Video & 20 videos & 20 hours & 30 & EEG, ECG, EDA, temp., resp., face and body video, gaze, audio & First-person & Discrete (9-step) & Arousal, valence, dominance, emotional keywords, predictability  & self-report & 20 \\
        \hline
        DEAP \cite{koelstra2012deap} & No & Video & 40 videos & 40 mins & 32 & EEG, BVP, EDA, EMG, temp., resp., face video & First-person & Discrete (5-step) & Arousal, valence, dominance, liking, familiarity  & self-report & 40 \\
        \hline
        LIRIS-ACCEDE \cite{baveye2015liris} & No & Video & $9,800$ videos & 27 hours & N/A & N/A & First-person & Pairwise & Arousal, valence   & 1517(arousal) 2442(valence) & UNK \\
        \hline
        Aff-Wild \cite{zafeiriou2017affwild} & No & Video & 298 videos & 30 hours & 200 & N/A & Third-person & Continuous bounded & Arousal, valence   & 6-8 & 298 \\
        \hline
        AffectNet \cite{mollahosseini2017affectnet} & No & Image & $450,000$ images & N/A & N/A & N/A & Third-person & Continuous bounded, categorical & Arousal, valence, 8 emotion categories  & 12 & $137,500$ \\
        \hline
        Sonancia\cite{lopes2017modelling} & No & Audio & 1280 sounds & N/A & N/A & N/A & First-person & Pairwise & Arousal, valence, tension   & UNK & ~10 \\
        \hline
        SEWA DB \cite{kossaifi2019sewa} & Yes & Video & 4 videos 1 task & 27 hours 17 hours & 398 & Facial landmarks, FAU, hand and head gestures & Third-person & Continuous bounded & Arousal, valence (dis)liking intensity, agreement, mimicry & 5 & 90 \\
        \hline
        RELOCA \cite{ringeval2013reloca} & Yes & Video & 1 task & 4 hours & 46 & ECG, EDA, face video,  audio & Third-person & Continuous bounded & Arousal, valence & 6 & 23 \\
        \hline
        GAME-ON \cite{maman2020game} & Yes & Social game & 1 game & 11.5 hours & 51 & Video, audio, and motion capture data & First-person & Discrete (5--9-step) & Emotions, cohesion, warmth, competence, competitivity, leadership, and motivation & self-report & 5 \\
        \hline
        MUMBAI \cite{doyran2021mumbai} & Yes & Board-game & 6 games & 46 hours & 58 & Gameplay, facial video, and facial action units  & First-person and Third-person & Discrete labels & Valence, attention, gameplay experience, personality & 56 (Third-person) 58 (First-person) & 6 \\
        \hline
        MazeBall \cite{yannakakis2010mazeball} & Yes & Videogame & 1 game & N/A & 36 & BVP(HRV), EDA, game telemetry & First-person & Pairwise & Fun, challenge, frustration, anxiety, boredom, excitement, relaxation & self-report & 1 \\
        \hline
        PED \cite{karpouzis2015platformer} & Yes & Videogame & 1 game & 6 hours & 58 & Gaze, head position, game telemetry & First-person & Discrete  (5-step), pairwise & Engagement, frustration, challenge  & self-report & 1 \\
        \hline
        FUNii \cite{beaudoin2019funii} & Yes & Videogame & 2 games & N/A & 190 & ECG, EDA, gaze and head position, controller input & First-person & Continuous, discrete & Fun (cont.), fun, difficulty, workload, immersion, UX & self-report & 2 \\
        \hline\hline
        \textbf{AGAIN} & \textbf{Yes} & \textbf{Videogame} & \textbf{9 games} & \textbf{37 hours} & \textbf{124} & \textbf{Game video, game telemetry} & \textbf{First-person} & \textbf{Continuous unbounded} & \textbf{Arousal} & \textbf{self-report} & \textbf{9} \\
        \hline
    \end{tabular}}
\end{sidewaystable*}

The availability of large-scale corpora comprising affect manifestations that are elicited through appropriate stimuli is a necessity for affect modelling. Creating datasets that are annotated with reliable affect information is, therefore, instrumental to the field of AC at large. In this section we review representative affective corpora that rely on audiovisual elicitors and discuss the contribution of AGAIN to the current list of datasets that are enriched with affect labels.
Table \ref{tab:datsets} presents the outcome of our survey\footnote{N/A indicates where the category is ``not-applicable'' (e.g. there are no participants when third-party videos are used) and UNK indicates if an attribute is ``unknown''.}. We follow a \emph{systematic} approach for reviewing the state of the art in affect corpora and examine the following factors that distinguish the surveyed datasets: the \emph{interactivity} and the \emph{type} of the provided elicitors, the number of possible elicitor \emph{items}, and the overall size of the available \emph{video} database (see second to fifth column of Table \ref{tab:datsets}), the \emph{number} of participants and their recorded \emph{modalities} (see columns six and seven of Table \ref{tab:datsets}), the annotation protocol in terms of \emph{perspective} and \emph{type} of annotation (see columns eight and nine of Table \ref{tab:datsets}), the affect \emph{labels} (see column ten of Table \ref{tab:datsets}), and finally the number of \emph{annotators} (if different from the number of participants and/or not self-reported) and number of \emph{tasks} each annotator had to complete (see the last two columns of Table \ref{tab:datsets}). 

It is apparent from Table \ref{tab:datsets} that affective datasets have gradually---over the last decade or so---drifted away from traditional induced elicitation and posed expressions, and instead turned towards soliciting spontaneous emotion manifestations. 
New datasets have been focusing on elicitation through naturalistic expressions. While some of these datasets use their own elicitors, many rely on popular media, using video clips and still images from music videos and movies \cite{koelstra2012deap,zafeiriou2017affwild,mollahosseini2017affectnet}. This method has proved to be reliable, cost-effective, and easy to set up, which subsequently led to a widespread adoption in the field. Compared to staged videos and images, interactive elicitors provide more organic stimuli. While reactions to non-interactive media can produce spontaneous expressions, interactive elicitation can increase the participants' involvement with the elicitor and reveal emotional reactions that might be hard to elicit with pre-recorded videos and images alone. Subsequently, there has been a growing body of research dedicated to enrich the set of affective corpora with interactive elicitors. These datasets use a wide-range of methods including dyadic tasks \cite{ringeval2013reloca,kossaifi2019sewa}, group tasks\cite{maman2020game}, board games\cite{doyran2021mumbai}, and video games \cite{karpouzis2015platformer,beaudoin2019funii}. These interactive tasks provide a more complex and multifaceted affective stimulus, while organically structuring the participants' experience. 

Most traditional affective computing databases surveyed capture affective dimensions such as \emph{arousal} and \emph{valence}, with some datasets offering labels for additional dimensions---such as \emph{dominance}---and categorical labels (see the \emph{annotation/labels} column in Table \ref{tab:datsets}). However, datasets using interactive elicitors tend to have a wider focus. While some of these datasets collect affective and emotional labels, their primary focus is task-related emotional outcomes. In datasets focusing on gameplay, this generally includes gameplay experience \cite{doyran2021mumbai} and other game-related outcomes--such as frustration, perceived challenge \cite{yannakakis2010mazeball,karpouzis2015platformer}, engagement \cite{karpouzis2015platformer}, and fun \cite{yannakakis2010mazeball,beaudoin2019funii}. Studies that use the same affective labels are easier to compare and their lessons easier to transfer to new data than studies using a more diverse set of labels (i.e. fun, engagement, challenge, etc.). On one hand, the mapping between labels such as ``fun'' and ``challenge'' can be uncertain; and on the other hand, it can be difficult to reliably translate outcomes such as ``engagement'' to more traditional affective computing concepts such as arousal or valence.

The affective datasets we survey appear to be rather split in terms of annotation type used. While some (e.g. DEAP \cite{koelstra2012deap}, MANHOB-HCI \cite{lichtenauer2011mahnob}) opt for self-reporting (first-person annotation), many databases (e.g. RELOCA \cite{ringeval2013reloca}, SEWA \cite{kossaifi2019sewa}) use only a few expert annotators in a third-person manner. There is a clear trade-off between these approaches. First-person annotations (i.e. self-reported labels) are ideal for capturing the subjective appraisal of emotional content, while third-person annotations are better at labelling emotion manifestation through inter-rater agreement \cite{afzal2011natural}. Interestingly, most datasets using an interactive elicitor also opt for first-person annotation through self-reports. A possible explanation is the higher degree of participant involvement, which makes the experience more unique to the participant. Such a highly subjective experience is better captured via first-person reporting. There is a definite trade-off, however, between interactive and non-interactive elicitation. Using multimedia clips for elicitation offers a cost-effective solution, which leads to large and varied datasets. On the other hand, interactive elicitors can stimulate emergent emotions more naturally.

Datasets using non-interactive elicitors are generally larger, while the cost associated with using interactive elicitation limits these datasets. Table~\ref{tab:datsets} shows that interactive datasets often focus on fewer elicitor items. However, this lack of variety is often offset by the different ways participants can interact with these elicitors, producing more diverse data. Despite this diversity, there are not many datasets that feature multiple different interactive elicitors. The handful of examples either combine dissimilar datasets \cite{camilleri2017towards}---which comes with its own challenges in reconciling different data formats---or use a small set of very similar elicitors---e.g. the FUNii dataset \cite{beaudoin2019funii} features two similar games from the same franchise. There are some exceptions: for instance, the MUMBAI dataset \cite{doyran2021mumbai} features six board games, although the data here is not self-reported but labelled by third-person annotators.

\begin{figure*}[t]
\centering
\subfloat[TinyCars]{\includegraphics[width=0.32\textwidth]{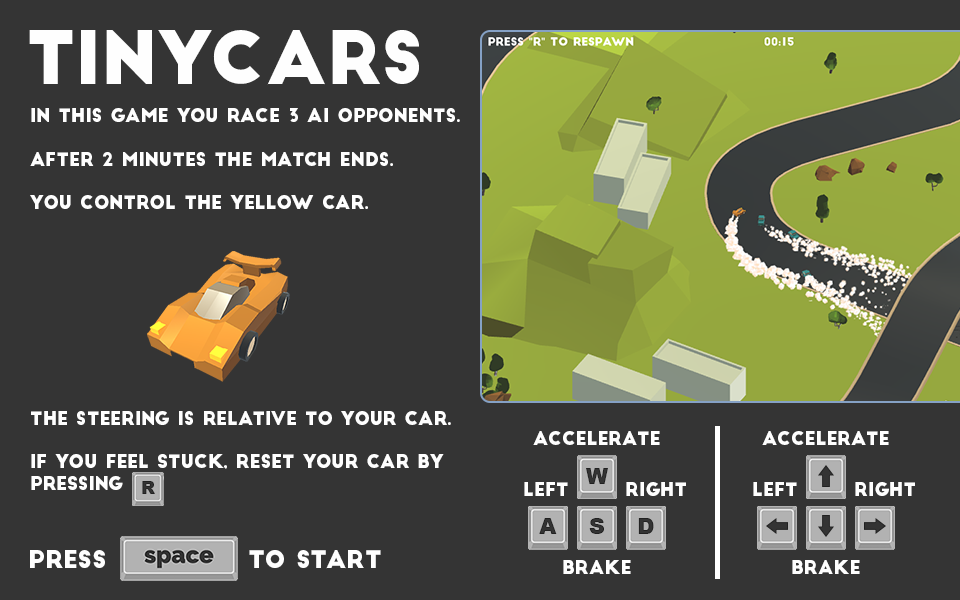}\label{fig:game:tiny}} \quad
\subfloat[Solid]{\includegraphics[width=0.32\textwidth]{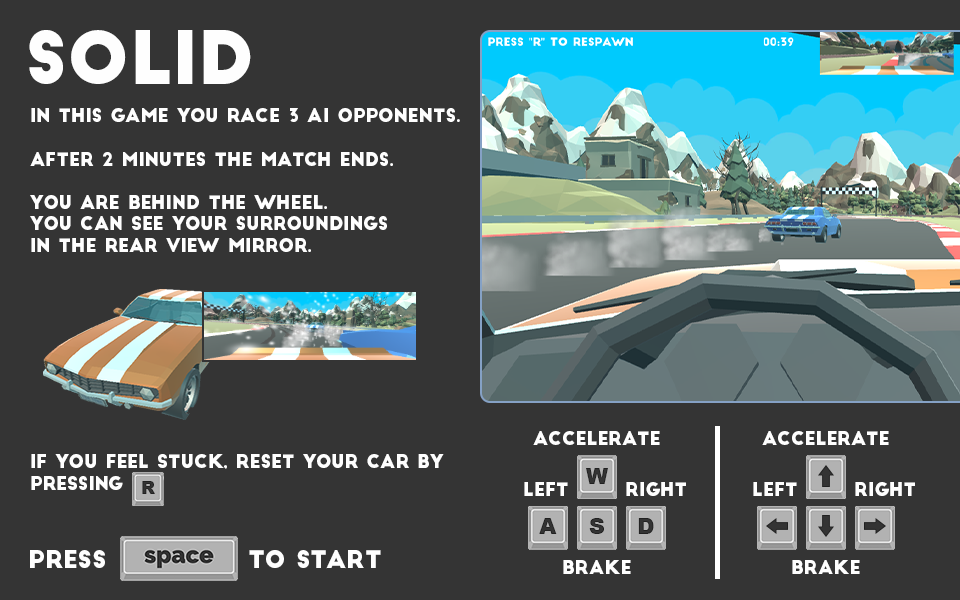}\label{fig:game:solid}} \quad
\subfloat[ApexSpeed]{\includegraphics[width=0.32\textwidth]{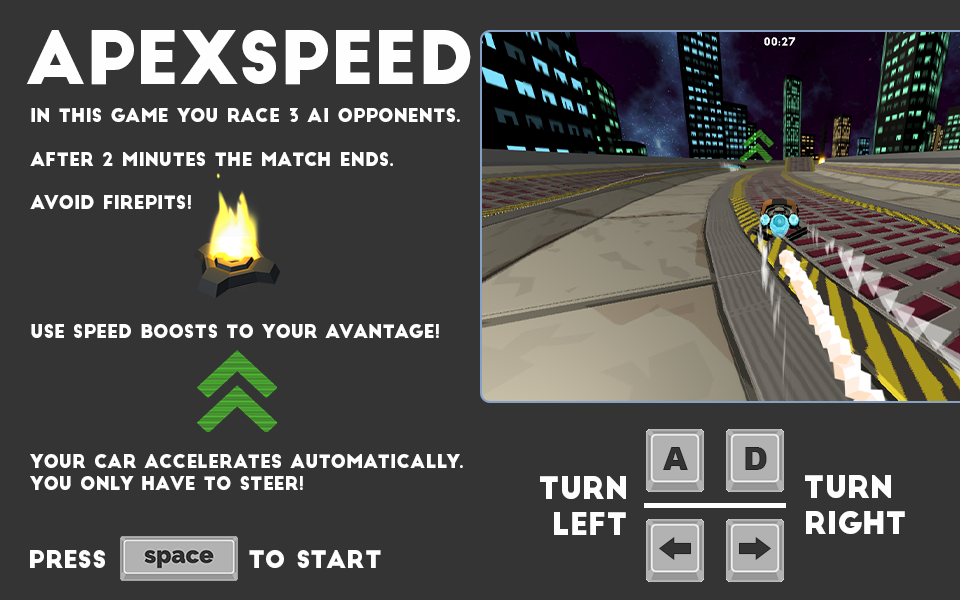}\label{fig:game:apex}} \\
\subfloat[Heist!]{\includegraphics[width=0.32\textwidth]{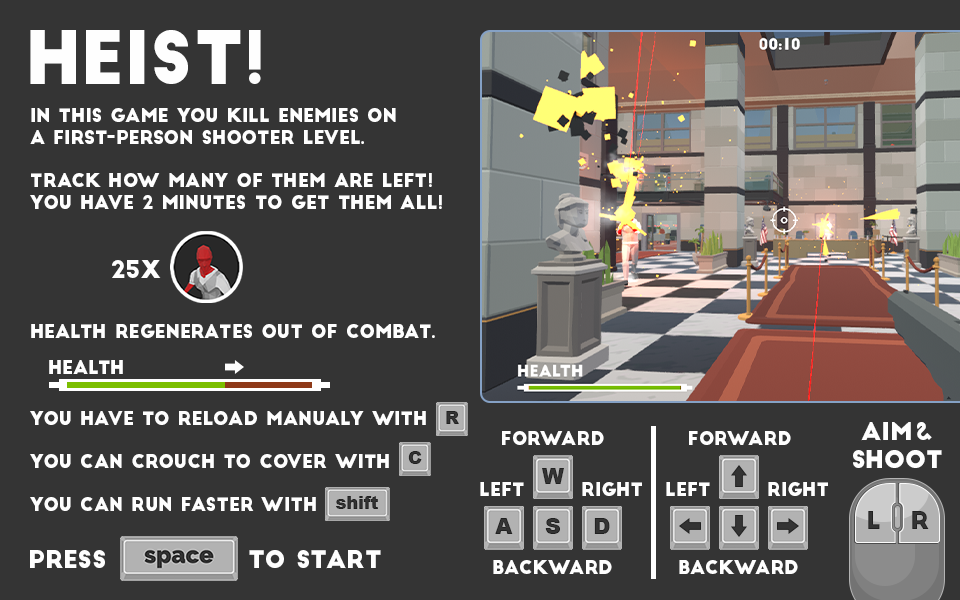}\label{fig:game:fps}} \quad
\subfloat[TopDown]{\includegraphics[width=0.32\textwidth]{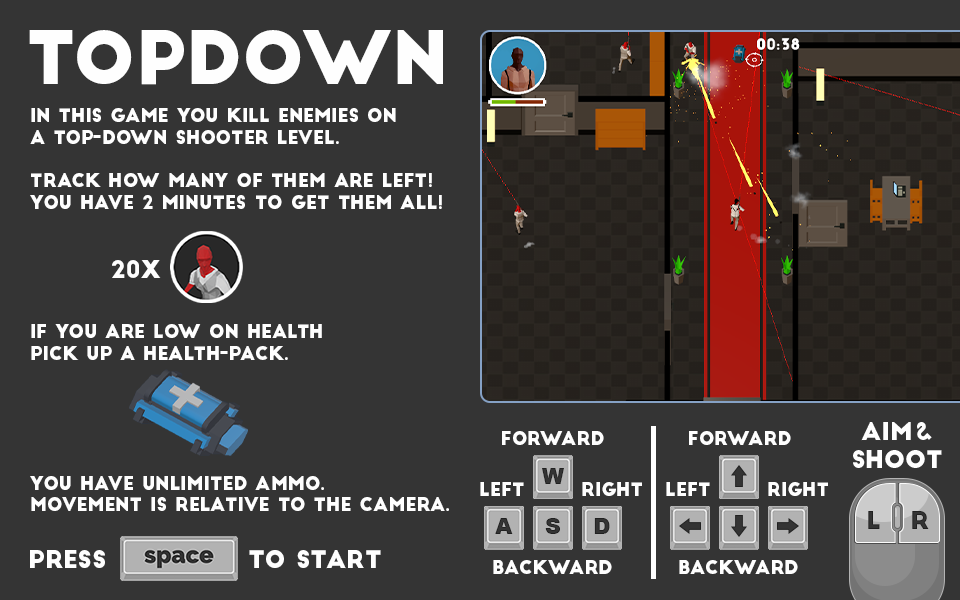}\label{fig:game:topdown}} \quad
\subfloat[Shootout]{\includegraphics[width=0.32\textwidth]{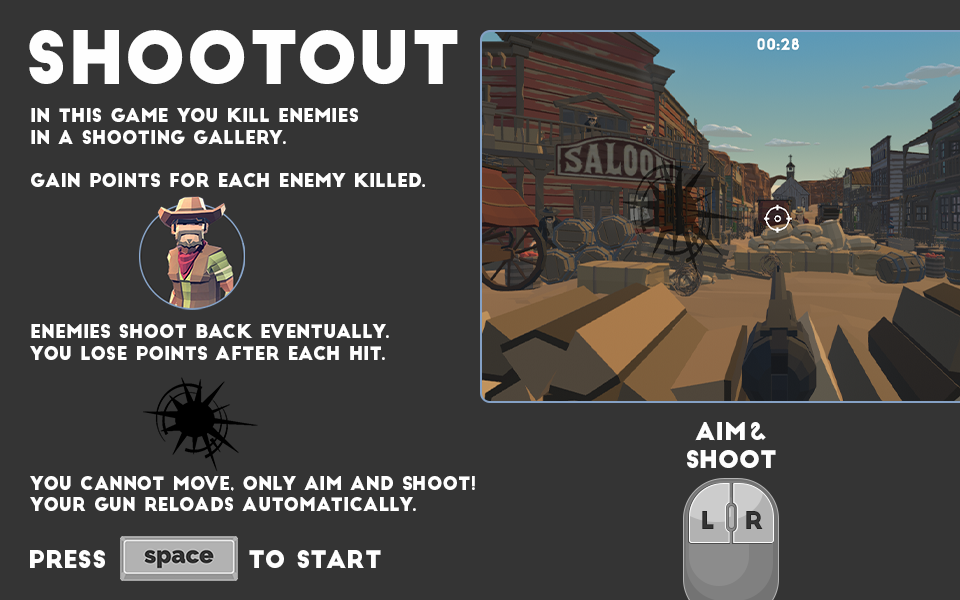}\label{fig:game:gallery}} \\
\subfloat[Endless]{\includegraphics[width=0.32\textwidth]{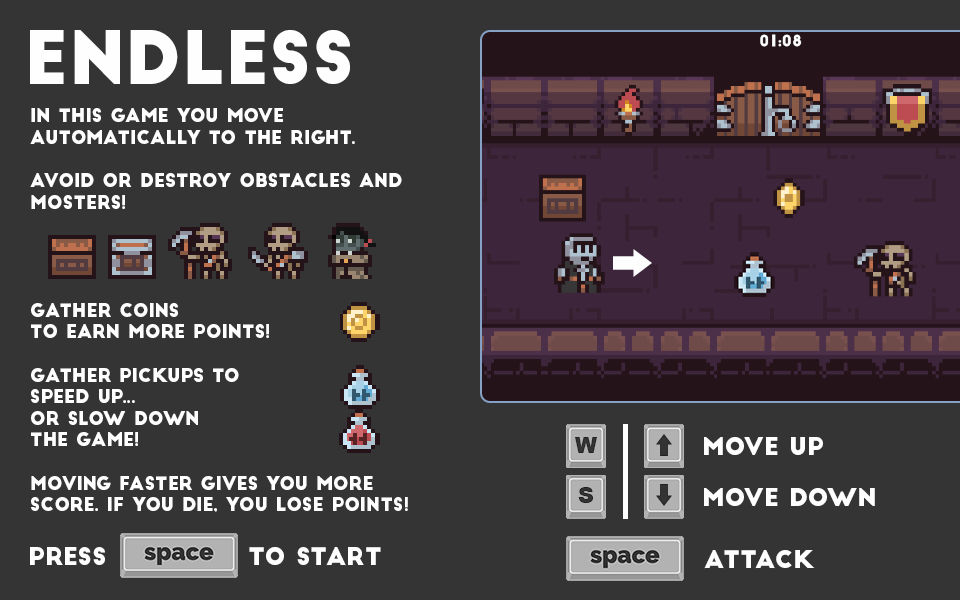}\label{fig:game:endless}} \quad
\subfloat[Pirates!]{\includegraphics[width=0.32\textwidth]{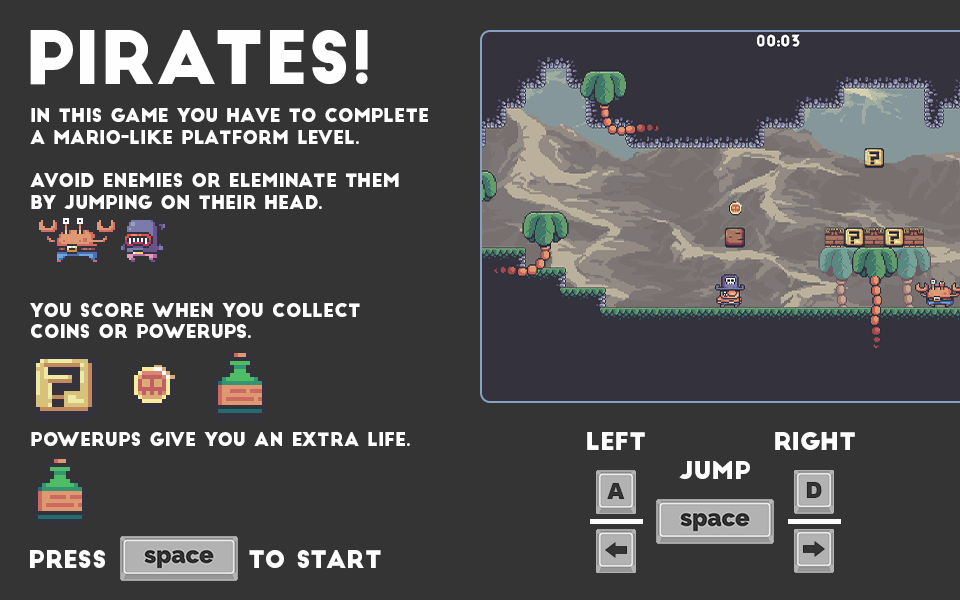}\label{fig:game:platform}} \quad
\subfloat[Run'N'Gun!]{\includegraphics[width=0.32\textwidth]{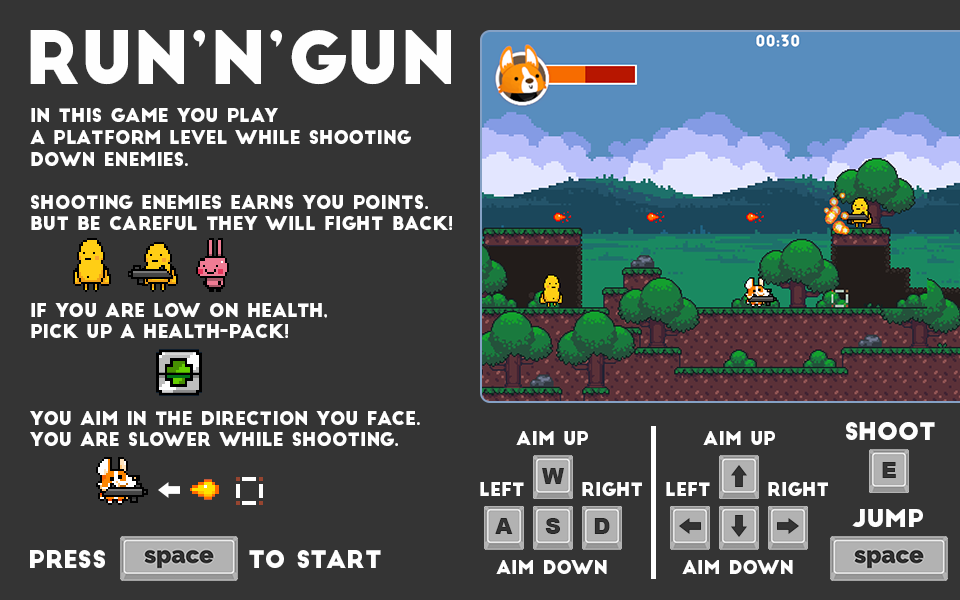}\label{fig:game:gun}}
\caption{Start screens of the nine games included in the AGAIN dataset, showing the game's rules and players' controls.}\label{fig:games}
\end{figure*}

AGAIN addresses the aforementioned limitations by offering a large-scale corpus that is based on a set of dissimilar interactive affect elicitors that are annotated through a first-person protocol. While the dataset at the time of writing is limited to 9 games and their annotated arousal, the dataset is planned to be augmented through more affective dimensions and enriched through more games. The resulting dataset leverages the strength of active emotion elicitation while producing data in volumes comparable to databases featuring non-interactive affect stimuli. Moreover AGAIN provides a diverse database for general player affect modelling research that is not possible within any of the existing corpora.

We position AGAIN at the intersection of traditional affective computing corpora and datasets with a focus on interactive emotion elicitation. By focusing on a core affect dimension (i.e. arousal) instead of task-related complex emotional outcomes, we aim to make the dataset more relevant to traditional AC research. We argue that the use of video games as interactive elicitors combined with traditional affective labels can also help bridge the gap between AC and games user research. As games are highly interactive media, the captured data and annotations encode not merely player affect but also behaviour and game context. We focus on self-reported labels to better capture the subjective intricacies of gameplay. Finally, we choose to record continuous unbounded traces of arousal using \emph{RankTrace} \cite{lopes2017ranktrace} via the PAGAN online annotation framework \cite{melhart2019pagan}. Such traces can be processed and machine learned in a number of ways including regression, classification and relational learning \cite{yannakakis2018ordinal}.

\section{Games \label{sec:games}}

Nine games, across three different genres, were designed and developed as affect elicitors specifically for the AGAIN dataset. We put careful consideration to create software which is aesthetically pleasing, representative of popular sub-genres of games, can be understood immediately with a basic level of game literacy \cite{buckingham2007game}, and produces a coherent and consistent dataset without the need of heavy pre-processing. The game genres were selected (racing, shooters, platformers) because they represent a good cross-section of the game genres \cite{vargas2020genre} and are among the most popular among gamers \cite{yannakakis2018artificial,sevin2020video}, but also because they have simple enough controls and clear mechanics so that players can pick them up quickly. Opposed to other genres, such as role playing or strategy games, that require longer time investment and players to learn the specific mechanics, strategies and synergies, the games in the dataset relied on fast-paced genres and popular tropes to communicate the game rules as fast as possible. Specific games were designed under each genre are representative of the genre.

\subsection{Racing}
The racing genre is characterised by fast-paced driving against a number of opponents in a given track. The dynamics of the experience is partly dictated by the limited interaction with opponent vehicles (e.g. pushing into each other) but mainly defined by the track itself. In the AGAIN database, three games are representing distinct sub-genres of racing games. \textbf{TinyCars} is an arcade-style racer with an isometric view (see Fig~\ref{fig:game:tiny}). Its controls are the hardest to master due to the drifting of the player's car. \textbf{Solid} is a more traditional rally game, with more realistic handling (see Fig~\ref{fig:game:solid}). As the player sees the track from the driver's seat, adapting to the turns of the track is more challenging. \textbf{ApexSpeed} is a speed-racer type game, with minimalist controls (see Fig~\ref{fig:game:apex}). While the player only has to change lanes (the vehicle accelerates and follows the track automatically), the game has a faster pace than other racing games and additional elements are complicating the track (i.e. speed boost platforms and obstacles).

\subsection{Shooter}
The shooter genre focuses on eliminating opponents using projectile weapons. The gameplay dynamic of these games builds on hand-eye coordination and it is characterised by periods of suspense and periods of engagement with the in-game opponents. Shooter games in the AGAIN dataset provide examples of different shooter sub-genres. \textbf{Heist!} is a typical first-person shooter game with similar mechanics to modern shooters (see Fig~\ref{fig:game:fps}). Because the player has to wait for their health to regenerate, the play experience is broken up into smaller engagements. In contrast, \textbf{TopDown} has a top-down view, an automatic weapon, and health pickups (see Fig~\ref{fig:game:topdown}). This provides a more action-packed environment as the player is not encouraged to stop if they are low on health. These two games also feature less linear maps compared to other games in the dataset. \textbf{Shootout} on the other hand does not feature traversal at all. In this game the player can only aim and shoot as the screen is filled with more and more enemies (see Fig~\ref{fig:game:gallery}). This dynamic of even-increasing intensity is typical of arcade-style games (including shooters).

\subsection{Platformer}
The platformer genre focuses on traversal and often requires precision and dexterity. The platformer games featured in the AGAIN dataset are the most diverse set of games. \textbf{Endless} is an endless-runner, a popular mobile-game genre. In these games, the player moves forward automatically at an ever-increasing pace while they have to attack or dodge incoming obstacles (see Fig~\ref{fig:game:endless}). Subsequently, Endless is one of the most frantic games in the dataset. \textbf{Pirates!} is a classical platformer, akin to Super Mario Bros (Nintendo, 1985) (see Fig~\ref{fig:game:platform}). This game has a more relaxed pace as the gameplay is focused on light platform puzzles and simple traversal. Finally, \textbf{Run'N'Gun} is a shoot-em up game, which has the characteristics of both a platformer and shooter game (see Fig~\ref{fig:game:gun}). Thanks to the shooting mechanics, number of enemies, and enemy projectiles, this game has a more intense gameplay loop compared to the other platformer games.

\section{AGAIN Dataset\label{sec:annotation}} 
Games in the AGAIN dataset were built for the WebGL platform and are played in a web-browser. The games were integrated into the PAGAN annotation platform \cite{melhart2019pagan}, which allowed the large-scale crowd-sourcing of both the game playing and annotation tasks.

\subsection{Protocol}\label{sec:annotation_protocol}

\begin{figure}[tb]
    \centering
    \includegraphics[width=1.0\linewidth]
    {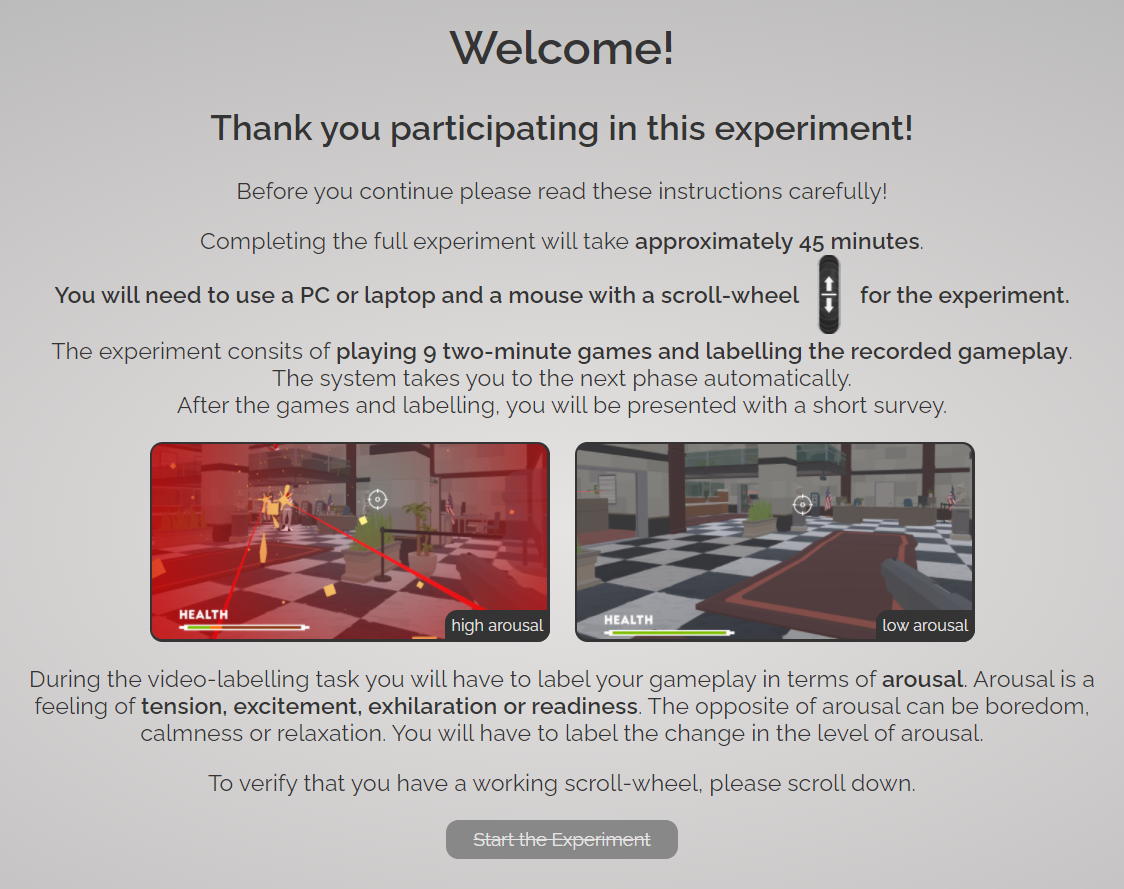}
    \caption{Introduction screen of the experiment.}
    \label{fig:pagan:intro}
\end{figure}

The collection procedure took anywhere between 45 to 55 minutes and followed by a stimulated recall protocol \cite{lankoski2015game}. Participants were invited through Amazon's Mechanical Turk service\footnote{\url{https://requester.mturk.com/}} and were compensated with 10 USD for their time. The only criterion for participation was prior purchase of video games, in order to filter out potential subjects who might not have the game literacy required to play the games. Participants were greeted with an introduction screen (see Fig.~\ref{fig:pagan:intro}), which informed them about the overall task and explained arousal as \emph{a feeling of tension, excitement, exhilaration or readiness} and \emph{the opposite of boredom, calmness or relaxation}. The experiment consisted of 9 rounds, each round consisting of 2 minutes of game-play followed by 2 minutes of annotation. Due to the high cognitive load of video game play, the annotations could not be collected at the same time as the game play telemetry. To mitigate this issue, a stimulated recall technique was used. Participants' gameplay was captured and played back to them during the annotation process. The collection procedure was set up in an iterative manner with participants playing for 2 minutes, then annotating their gameplay video for 2 minutes. The order of the games was randomised and this procedure was repeated until all games were played and annotated. After the experiment, participants filled in a simple exit-survey recording their biographical data and gaming habits. 

\subsection{Participants}\label{sec:annotation_collection}

Through the procedure presented in Section \ref{sec:annotation_protocol}, we collected data from 124 participants\footnote{While 169 participants completed the data collection process, 45 participants were omitted as their experiments were incomplete (i.e. no video or annotation data) due to software or hardware error.} which include $1,116$ gameplay sessions (124 sessions per game) with detailed telemetry and over 37 hours of gameplay videos. Out of the 124 participants, one identified as non-binary, 43 as female, and 80 as male. Participants' age varied between 19 and 55 years old (average of 33). Most participants were from the USA (82\%); the remaining 22 participants came from Brazil (10 participants), Italy (3), Canada (2), India (2), Czech Republic (1), Germany (1), and Romania (1). Most participants identified as casual gamers (57\%) or hard-core gamers (36\%). Reflectively, the majority of participants (87\%) were playing daily or weekly. All participants had either a PC or a gaming console or both, with the most popular platform being PC. Participants played very diverse games in their free time across different genres: from casual games through platformers, sports simulators, shooters, to role-playing games. The anonymised demographic data is included in the dataset.

\subsection{Game Footage Videos}\label{sec:videos}

For realising first person annotation, the gameplay footage of players had to be recorded and annotated by the players themselves. As a result the raw AGAIN dataset features $1,116$ videos of around 2 mins each (i.e. over 37 hours of game footage). The video database contains more than $3\times10^6$ frames of video, which are recorded at $24$ FPS and have a resolution of $960\times600$ pixels. Such data can enable future research that employs computer vision and deep-learning to directly map pixels to emotions \cite{makantasis2019pixels}. Previous studies have shown promise in using general-purpose representations such as pixel data from game footage \cite{makantasis2021pixels}, and utilising audiovisual data for learning through privileged information \cite{makantasis2021privileged}.

\subsection{Game Context Features}\label{sec:features}

\begin{table}[!tb]
    \centering
    \caption{Number of features extracted per game}
    \label{tab:num_features}
    \begin{tabular}{|l|l|l|c|}
        \hline
        \textbf{Genre} & \textbf{Sub-Genre} & \textbf{Title} & \textbf{\# Features}\\
        \hline
        \hline
        Racing & Arcade-Racing & TinyCars & 33\\
           & Rally & Solid & 34\\
          & Speed-Racer & ApexSpeed & 34\\
        \hline  
        Shooter & First-Person Shooter & Heist! & 37\\
          & Top-down Shooter & TopDown & 38\\
          & Arcade-Shooter & Shootout & 23\\
        \hline  
        Platform & Endless Runner & Endless & 33\\
           & Mario-Clone & Pirates! & 39\\
          & Shoot'Em'Up & Run'N'Gun & 47\\
        \hline  
    \end{tabular}
\end{table}

In addition to the raw video game footage, AGAIN features a number of hand-crafted attributes for each game. Inspired by advances in machine learning with privileged information \cite{vapnik2009new,makantasis2021privileged} we view telemetry data as \emph{privileged information} and we include such ad-hoc features in the dataset. Privileged information here means information that pertains to an experience but not readily available to an observer. This kind of information generally encodes domain specific or hard-to-attain data. The associated cost of learning the information makes the data valuable in building expert systems, but poses some limitation to data-hungry machine learning approaches. When it comes to video games, privileged information can include player physiology or game telemetry based on expert heuristics \cite{makantasis2021privileged}. Fusing gameplay features with other user modalities has also been a dominant practice in game-based affective computing \cite{martinez2014deep,martinez2014don}. The game context features described in this section are considered in the preliminary data analysis of the dataset in Section \ref{sec:annotation}.

\begin{table}[!tb]
    \centering
    \caption{The general gameplay features of AGAIN}
    \begin{tabular}{|l|l|}
        \hline
        \textbf{feature} & \textbf{description}\\
        \hline
        \hline
        \texttt{time\_passed} & time counted from the start of the recording \\
        \texttt{score} & player score \\
        \texttt{input\_intensity} & number of keypresses \\
        \texttt{input\_diversity} & number of unique keypresses \\
        \texttt{idle\_time} & percentage of time spent without input \\
        \texttt{activity} & inverse of \texttt{idle\_time} \\
        \texttt{movement} & distance travelled + reticle moved (in shooters) \\
        \texttt{bot\_count} & number of bots visible \\
        \texttt{bot\_movement} & bot distance travelled \\
        \texttt{bot\_diversity} & number of unique bots visible \\
        \texttt{object\_intensity} & number of objects of interest \\
        \texttt{object\_diversity} & number of unique objects \\
        \texttt{event\_intensity} & number of events \\
        \texttt{event\_diversity} & number of unique events \\
        \hline
    \end{tabular}
    \label{tab:general_features}
\end{table}

All AGAIN games implement the same data-logging strategy and use a similar method for recording telemetry. Games within the same genre share the same feature labels. Not all features, however, have a qualitative meaning for all games within a genre---for instance, players move in \emph{Heist!} but are immobile in \emph{Shootout}. To ease the data collection and aggregation process, when features are absent from a game they are given values with zero-variance (zeroes or ones, depending on the feature). For example, a looping racetrack is only present in the \emph{Solid} game (see Figure \ref{fig:game:solid}), therefore the \texttt{visible\_loop\_count} feature is always zero in the other racing games. 

Table \ref{tab:num_features} shows the number of features we have extracted per game with the zero-variance features removed. Recorded game telemetry encodes \emph{control events} initiated by the player (e.g. \texttt{player\_steering}), \emph{player status} (e.g. \texttt{player\_health}), \emph{gameplay events} outside of the player's control (e.g. \texttt{bot\_aim\_at\_player}), \emph{bot status} (e.g. \texttt{bot\_offroad}), and the \emph{proximal} and \emph{general game context} (e.g \texttt{bot\_player\_distance} and \texttt{pickups\_visible}). Gameplay is recorded at approximately 4Hz (every 250ms). Due to limitations of the Unity engine and the WebGL format, the logging rate is not consistent. To mitigate this issue, the logging script aggregates multiple ticks of the engine's update loop and provides an average value. Due to this processing technique almost all events are represented by continuous values. For example, \texttt{pickups\_visible} can take float values under 1 when a pickup just became visible at the end of the given 250ms window. The only features which are represented by integer values are \texttt{player\_death} and \texttt{objects\_destroyed} because of their sparsity.

In addition to the features enumerated in Table \ref{tab:num_features}, the dataset includes 14 \emph{general gameplay features}. These general features are ad-hoc designed and derived from the game-specific events and are based on contemporary studies of general player modelling \cite{camilleri2017towards, bonometti2020theory}. Events which require expert evaluation of the game such as the \emph{goal-oriented} and \emph{goal-opposed} events of Camilleri et al. \cite{camilleri2017towards} are omitted from these general features of AGAIN, but may be considered as additional features. Table \ref{tab:general_features} lists these features alongside their explanation.

\subsection{Annotation}\label{sec:annotation_tools}

\begin{figure}[t]
    \centering
    \includegraphics[width=1.0\linewidth]
    {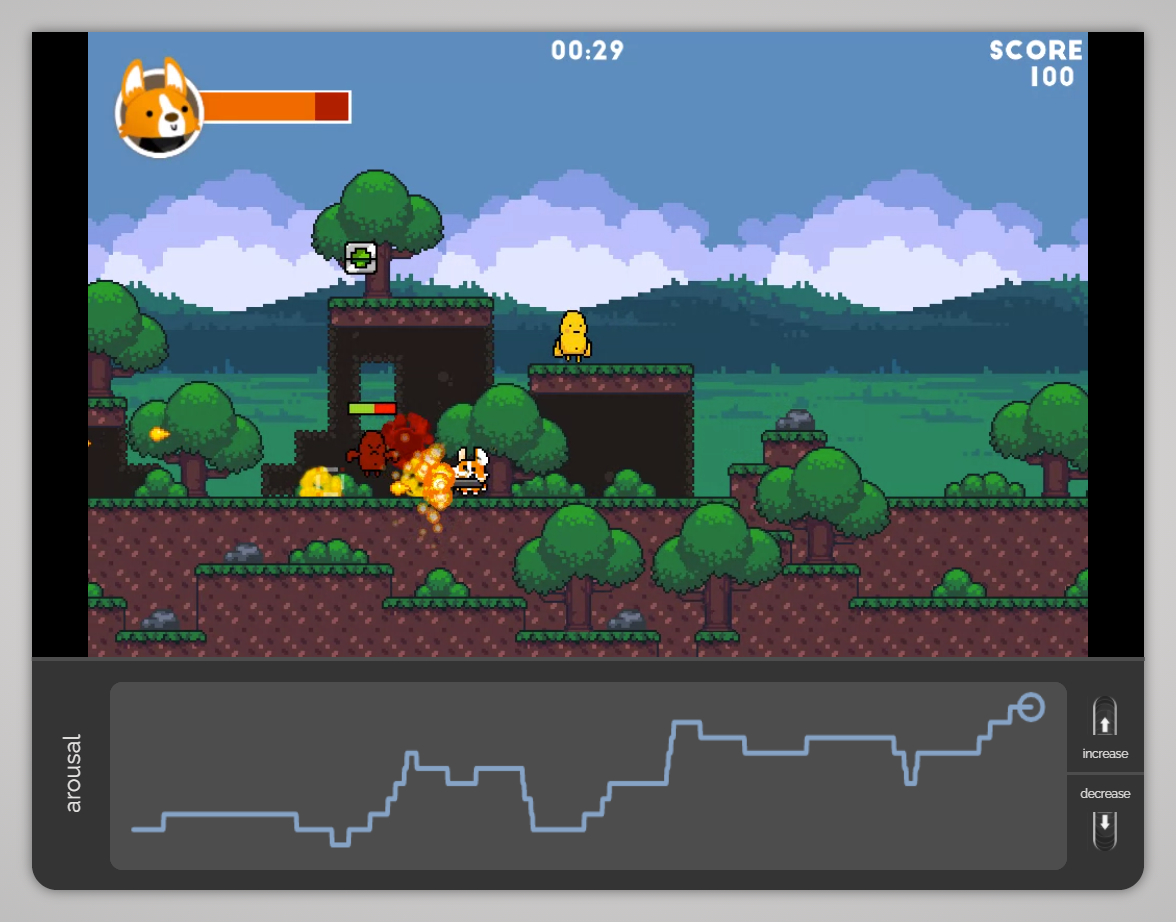}
    \caption{The PAGAN \emph{RankTrace} annotation interface. The gameplay video is played in the window above and the participant controls the annotation cursor (blue circle) below, drawing a visible annotation trace.}
    \label{fig:pagan:ranktrace}
\end{figure}

The annotation task was administered through the PAGAN platform \cite{melhart2019pagan}, using the \emph{RankTrace} annotation method \cite{lopes2017ranktrace}. PAGAN is an online annotation platform developed to be an easy-to-use software for crowdsourcing annotation tasks with a focus on \emph{one-dimensional time-continuous annotation} using three different methods. \emph{RankTrace} \cite{lopes2017ranktrace}, an ordinal annotation framework, \emph{GTrace} \cite{cowie2013gtrace}, a bounded annotation scale which gathers continuous data that can be converted to a Likert-like format, and \emph{BTrace}, which is a binary annotation tool for both time-continuous and discrete annotation, inspired by \emph{AffectRank} \cite{yannakakis2015grounding}. We have chosen \emph{RankTrace} as our annotation framework for this dataset.

\emph{RankTrace} allowed us to collect data in an unbounded fashion (see Fig. \ref{fig:pagan:ranktrace}). Due this collection method, the value range of the annotation is not bounded between 0 and 1, which can make it significantly harder to use the data for certain tasks---for example applying regression. However, this type of data is best interpreted as subjective, ordinal labels as it preserves the relative relationships between datapoints \cite{yannakakis2018ordinal},
and therefore is well suited for preference learning tasks. The unbounded trace means that users can always adjust their annotations higher or lower than previous values, which alleviates much of the guesswork compared to when users annotate on an absolute and objective scale \cite{martinez2014don}. The ordinal nature of the annotation follows the cognitive process of human evaluation, as it provides a trace which factors in habituation \cite{solomon1974opponent}, anchoring bias \cite{damasio1994descartes, seymour2008anchors} and recency-effects \cite{erk2003emotional}. To preserve this subjectivity encoded in the annotation, we apply data transformation (i.e. normalisation and pairwise transformation) based on individual sessions.

\subsection{Data Cleaning}\label{sec:annotation_cleanup}
    
\begin{figure}[!tb]
    \centering
    \includegraphics[width=1\linewidth]{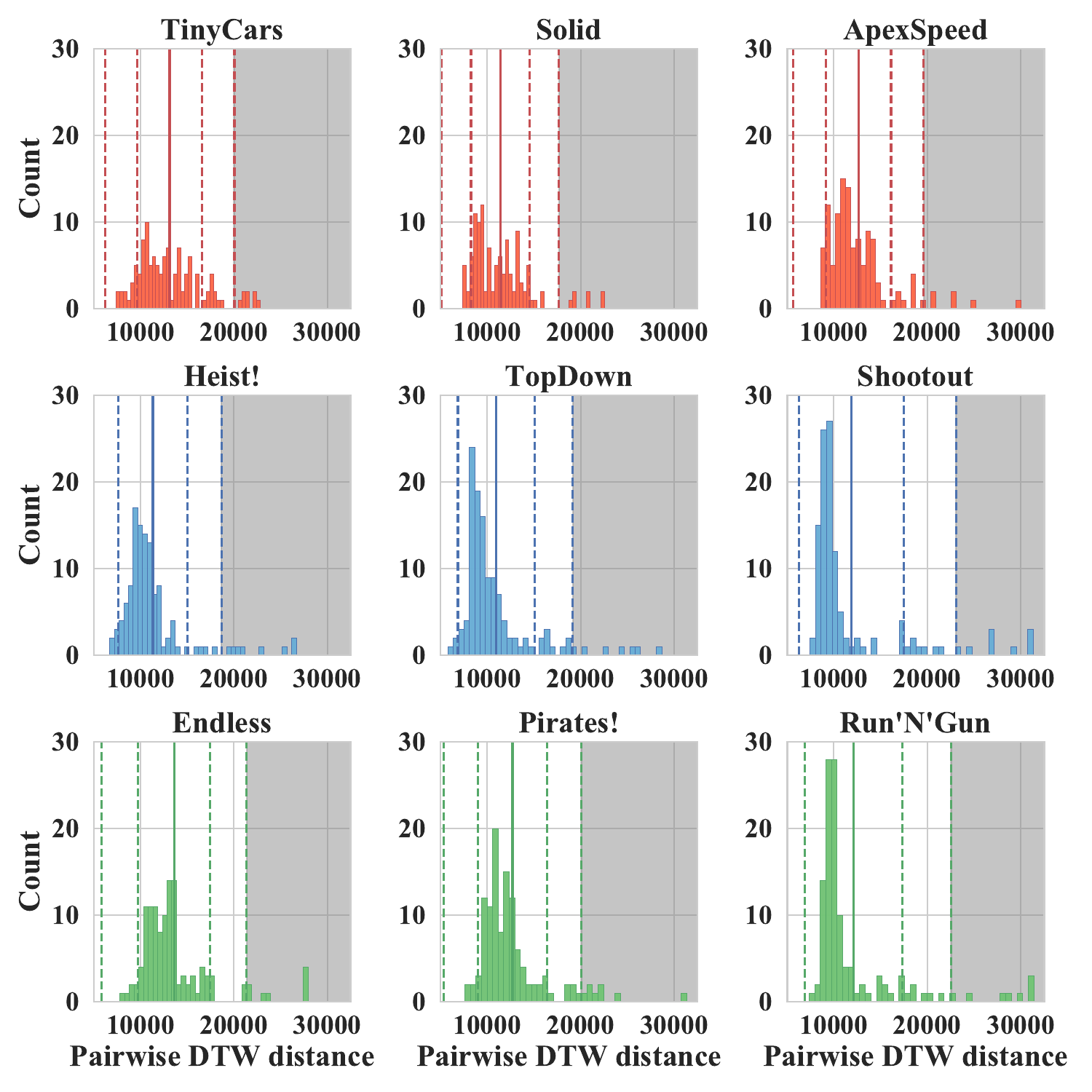}
    \caption{Distribution of summed cumulative DTW distance values of each session compared to every other session. The solid line shows the average score, while the dotted lines show the first and second standard deviation. Values in the grey field (right tail) are removed during data cleaning.} 
    \label{fig:dtw}
\end{figure}

To ease any subsequent analysis and future studies based on the dataset, in this section we propose a preprocessing pipeline which removes 10.8\% of the dataset as outliers. AGAIN contains both the raw and the cleaned data that result from the process outlined here. 

Since PAGAN only records annotations when there is a change in the signal and the Unity engine loop is affected by hardware performance, as a first step we resample the whole dataset at 4Hz to get a consistent signal. We remove duplicate values from the dataset, as well as sessions which are either too short (less than 1 minute) or too long (more than 3 minutes) due to software or technical errors during crowdsourcing. We also prune sessions which have less than 10 annotation points, assuming that the participant was unresponsive. This initial cleanup phase removes 24 sessions ($2.1\%$ of the data).

Inspired by Makantasis et al. \cite{makantasis2021pixels}, we apply Dynamic Time Warping (DTW) to clean the dataset of irregular annotations. DTW is used extensively in time-series analysis as a distance measure \cite{wang2013experimental,petitjean2016faster,li2020adaptively}. DTW is an elastic measure of distance between two signals, which can be of different length and sampled at different rates \cite{berndt1994using,wang2013experimental}. The \emph{DTW distance} is calculated based on the similarity matrix between two time-series, where every point of the two sequences is matched to each other---with one-to-many mapping where necessary \cite{berndt1994using}. While in signal processing DTW is often applied to synchronise different signals, the cumulative distance---calculated when finding the warping path between signals---provides a reliable similarity measure between the dynamics of the time-series in question \cite{wang2013experimental,petitjean2016faster}. 
We apply the cumulative DTW distance as a similarity measure between arousal traces, in order to remove irregular annotation patterns; we do not transform any of the signals. It should be noted that the signals have been synchronised and resampled, and thus the length and frequency of all annotation traces is the same.

\begin{table}[!tb]
\centering
\caption{Preliminary analysis of the clean AGAIN dataset. The table lists the number of game sessions and their corresponding data points on a frame-by-frame basis (250 ms). The table also lists the number of 3s time windows within which the arousal value increases ($\uparrow$), decreases ($\downarrow$) or stays stable within a 10\% threshold bound (---).}
\begin{tabular}{|l|c|c|c|c|c|}
\cline{4-6}
\multicolumn{3}{c}{} & \multicolumn{3}{|c|}{\textbf{Arousal (3 s interval)}} \\
\hline
\textbf{Game} & \textbf{Sessions} & \textbf{Data $(\cdot10^3)$} & \textbf{$\uparrow$} & \textbf{$\downarrow$} & \textbf{---} \\
\hline\hline
TinyCars & 109 & 52.75 & 543 & 461 & 3386 \\
Solid & 109 & 53.42 & 613 & 492 & 3346 \\
ApexSpeed & 114 & 56.10 & 607 & 462 & 3581 \\
\hline
\textbf{Racing} & 332 & 162.27 & 1763 & 1415 & 10313 \\
\hline\hline
Heist! & 110 & 53.91 & 580 & 424 & 3479 \\
TopDown & 115 & 56.90 & 650 & 463 & 3614 \\
Shootout & 106 & 51.77 & 471 & 341 & 3496 \\
\hline
\textbf{Shooter} & 331 & 162.57 & 1701 & 1228 & 10589 \\
\hline\hline
Endless & 112 & 55.11 & 559 & 438 & 3595 \\
Pirates! & 110 & 52.26 & 625 & 534 & 3186 \\
Run'N'Gun & 110 & 54.97 & 618 & 431 & 3521 \\
\hline
\textbf{Platformer} & 332 & 162.34 & 1802 & 1403 & 10302 \\
\hline\hline
\textbf{Total} & 995 & 487.18 & 5266 & 4046 & 31204 \\
\hline
\end{tabular}
\label{tab:clean_data}
\end{table}

As a first step in the cleanup process, we calculate the cumulative DTW distance to an artificial flat baseline (arousal annotations at 0 in all time windows). The resulting score provides us with a similarity measure to an artificial session where the participant performed no annotation; this allows us to remove unresponsive outliers. We remove all sessions which fall more than two standard deviations closer to zero from the average cumulative distance (the left tail of the distribution). This step removes 28 additional sessions from the dataset ($2.5\%$). Since games in the dataset are quite short and players encounter similar situations, we assume that their experience would be somewhat similar. Therefore, we remove sessions where the annotation traces are too far from other traces in the dataset. To this end, we apply the cumulative DTW distance metric between each datapoint and sum up the resulting distances. This metric shows us the relative similarity of a session to every other session. We remove all sessions which fall more than two standard deviations away from the average summed cumulative distance (see Fig.~\ref{fig:dtw}). This step removes an additional 69 sessions ($6.2\%$). This last step removes annotations which are too dissimilar from the general trends of participants' annotations; we presume that either the annotation was improper or that this session's elicitor was somehow not in line with how other players played the same game. Observing outliers empirically affirms that most participants whose annotation traces were atypical encountered issues with game controls, experienced slow-down and other glitches, or in most cases annotated \emph{positive} and \emph{negative} events instead of high and low arousal.

At the end of the cleaning process, 121 sessions---including all data from 2 participants---are removed ($10.8\%$). Around 40\% of the outliers are removed due to inactivity or incompleteness, while the rest is held out due to unusual annotation patterns. The cleaning process proposed in this section is conservative due to the limitations of the online collection process, where there is less control over the quality of annotation. However, the raw dataset is also made available, which provides opportunities for different processing methods. The clean dataset consists of 122 participants and 995 sessions; details on the clean dataset are provided in Section \ref{sec:dataset}.

\section{AGAIN Analysis}\label{sec:dataset}

Following the cleanup process presented in Section \ref{sec:annotation_cleanup}, this Section performs a preliminary analysis of the clean version of the AGAIN dataset, focusing on patterns in the arousal annotations and the AGAIN game context features (see Section \ref{sec:dataset_trends}). Section \ref{sec:dataset_baseline} describes an initial set of affect modelling experiments with this dataset, serving as baseline for future studies. While some games receive more aggressive data cleaning than others (\emph{TinyCars, Solid,} and \emph{Shootout}), overall there is an even distribution of data and sessions across genres as shown in Table~\ref{tab:clean_data}.

\subsection{Trends in the Data}\label{sec:dataset_trends}

\begin{figure}[!tb]
    \centering
    \includegraphics[width=1\linewidth]{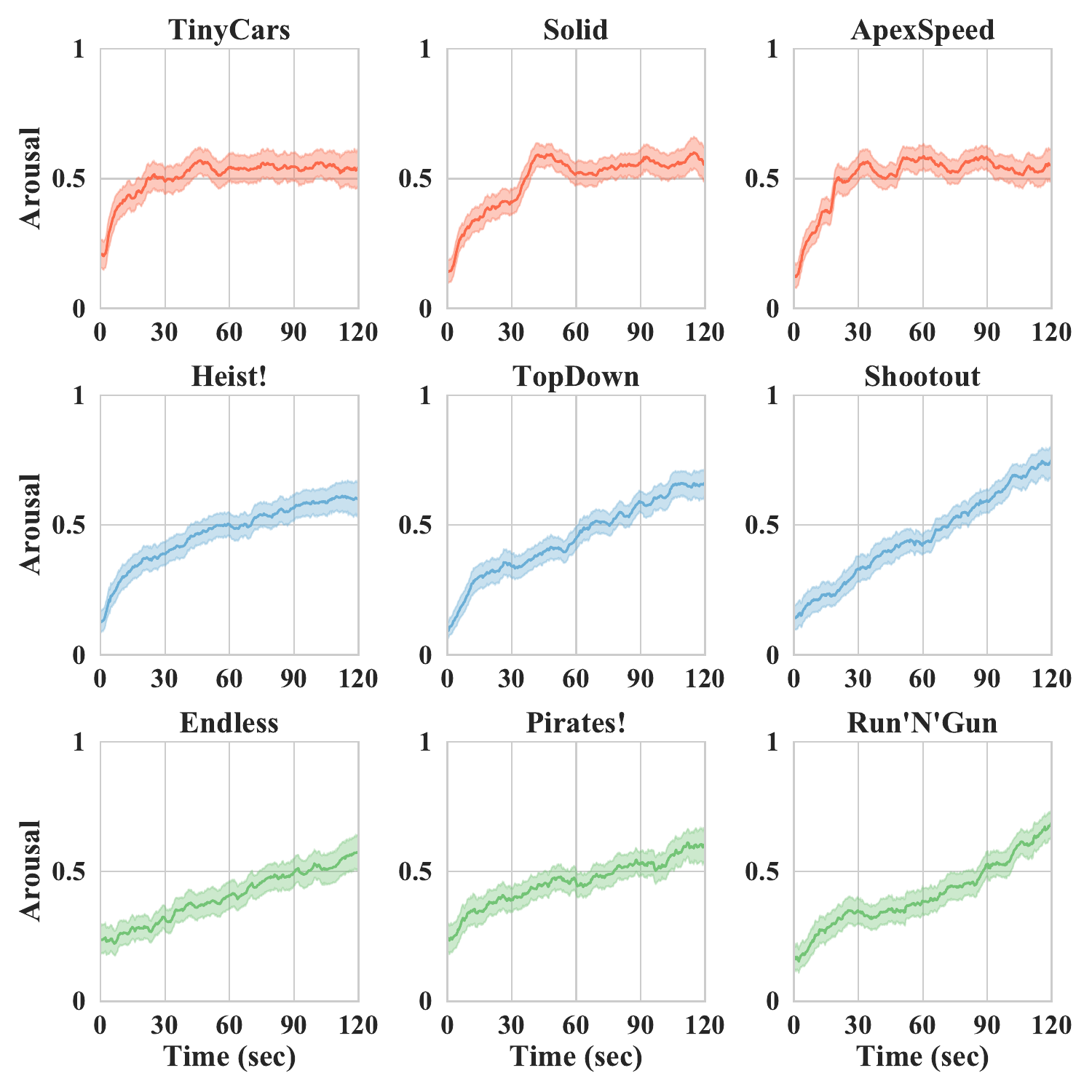}
    \caption{Average annotation traces (normalised per session) showing an increasing tendency. The coloured area around the mean depicts the 95\% confidence interval of the mean.}
    \label{fig:annotation}
\end{figure}

Figure \ref{fig:annotation} shows the average annotation trace as calculated by averaging values in time windows of 250 ms of all sessions' traces. The gameplay sessions have been normalised to show the relative trend in the data. It is evident that arousal annotation tends to have an upwards tendency. This is not surprising, as most games considered are action-oriented with an ever-increasing challenge; for instance, \emph{Endless} keeps increasing the speed of the game which evidently makes it both harder and more arousing as time passes. Racing games (top row of Figure \ref{fig:annotation}), on the other hand, tend to have arousal converging to a maximum mean value after the first 30 seconds. This is likely because the player is initially rushing to overtake the opponents' cars (players always start last); after this initial excitement the race becomes repetitive, with players trying to either maintain the lead or slowly catch up to the leader.

Observing the twelve general gameplay features shared across all nine games, one can detect some notable differences between games. In terms of the player's input (control), games with more complex interaction schemes appear to have higher input diversity and input intensity (see Table \ref{tab:general_features} for details on these features). Even accounting for the games' different control schemes (i.e. the number of controls the player has available), \emph{ApexSpeed, Shootout}, and \emph{Endless} have the lowest intensity (number of keypresses) and diversity (number of unique keypresses) while \emph{Pirates!} and \emph{TinyCars} have the highest diversity. This discrepancy could point to an easier control scheme for the former games, but it could also point to a more frantic and engaging interaction in the latter games. The idle time and activity features corroborate this observation, as racing games have less idle time without keypresses (since in two of the games the player needs to constantly press a button to move forward). In contrast, games where participants mainly reacted to stimuli (e.g. in \emph{Shootout} players react to opponents popping up and in \emph{Endless} players move only when a gap or obstacle is near) featured much higher idle times. In terms of other features, the number of bots (opponents) visible on the screen varied wildly between games, with \emph{Tiny Cars} and \emph{Shootout} having the highest number of visible enemies on average. Perhaps due to the many enemies present, \emph{Shootout} had the highest number of events (event intensity in Table \ref{tab:general_features}), while \emph{Solid} had the fewest events per time window.

\subsection{Preliminary Arousal Models}\label{sec:dataset_baseline}

In this section we provide an initial modelling approach for the AGAIN dataset, serving as a baseline study for future research with this dataset. As a preliminary step, we process the clean AGAIN dataset to predict arousal. 
To this end, we split the annotation traces into 3-second time windows---computing the mean of the window---and introduce a 1-second lag to the annotation trace. Our choice of time windows and lag is motivated by best practices established by a long line of prior research \cite{metallinou2013annotation,yannakakis2015grounding,lopes2017ranktrace,camilleri2017towards,melhart2019tom,melhart2021towards,pacheco2021trace}, as well as empirical results of studies into AC research design. It has been shown that a 3-second window size is well-suited to capture valence and arousal changes \cite{ayata2016emotion}. In their experiment on the DEAP dataset \cite{koelstra2012deap}, Ayata et al. have shown that affective data processed at this granularity leads to a higher model performance \cite{ayata2016emotion}. Mariooryad and Busso have shown that while an optimal input lag value can be found algorithmically, an ad-hoc value between 1 to 3 seconds gives a good approximation of human input lag in AC annotation tasks \cite{mariooryad2013analysis}. Here we chose a 1-second lag to conform to the aforementioned body of research. All features (including arousal values) are normalised on a per-session basis to a $[0,1]$ range. This means that feature values of 0 and 1 are indicating the minimum and maximum intensity of a given feature only within a session. This processing method gives weight to the relative dynamics of features instead of focusing on the absolute values.

While in the published dataset both clean and raw data is available for the application of different machine learning techniques, we treat arousal modelling in AGAIN as a \emph{preference learning task} \cite{furnkranz2011preference,yannakakis2017ordinal,yannakakis2018ordinal} and focus on predicting arousal \emph{change} from a 3-second time window to the next. We apply preference learning through a \emph{pairwise transformation}. During this transformation we observe consecutive datapoints within sessions in pairs and create a new representation of the dataset. By describing the difference between arousal values of time windows, this new representation reformulates the preference learning problem as binary classification (arousal increasing or decreasing). For every $(x_i, x_j) \in X$ pair of game data we observe the relationship of their affect output $(y_i, y_j) \in Y$. If $y_i$ is preferred to $y_j$, we can label the distance between the corresponding data points ($x_i-x_j$) and $1$. Conversely, we can label the reverse of this observation ($x_j-x_i$) as $0$. While either one of these observations is sufficient to describe the relationship between $x_i$ and $x_j$, by keeping both observations ($\lambda_{x_i-x_j} = 1$ and $\lambda_{x_j-x_i} = 0$), we can maintain a 50\% baseline accuracy in the post-transformation dataset independently of the trends in the dataset before the transformation. While this method creates redundancies in the training data, it mitigates some of the issues that arise from the strong temporal patterns discussed in Section~\ref{sec:dataset_trends}, as the algorithm is trained on both increasing and decreasing examples. To reduce experimental noise from trivial changes within the arousal trace, we omit all consecutive time windows between which the arousal change is less than 10\% of the total amplitude of the session's arousal value. While this 10\% threshold is based on prior experiments in similar problems \cite{lopes2017modelling, melhart2020study}, a more extensive analysis could explore the impact of the threshold value on prediction accuracy and the volume of data lost. 

As mentioned above, applying this pairwise transformation to consecutive time windows reformulates the preference learning paradigm as binary classification. To construct accessible and simple models of arousal, this initial study employs a \emph{Random Forest Classifier}. A Random Forest (RF) is an ensemble learning method, which operates by constructing a number of randomly initialised \emph{decision trees} and uses the \emph{mode} of their independent predictions as its output. Decision trees are simple learning algorithms, which operate through an acyclical network of nodes that split the decision process along smaller feature sets and model the prediction as a tree of decisions \cite{lewis2000introduction}. In this paper we are using the RF implementation in the \emph{Scikit-learn} Python library \cite{scikit-learn}. We initialise RFs with their default parameters. For controlling overfitting we set the number of estimators in the RF to $100$ and the maximum depth of each tree to $10$. This experimental setup is meant to provide a simple baseline prediction performance for the dataset, and thus we are not tuning the hyperparameters of the algorithm.

\begin{figure}[!tb]
    \centering
    \includegraphics[width=1\linewidth]{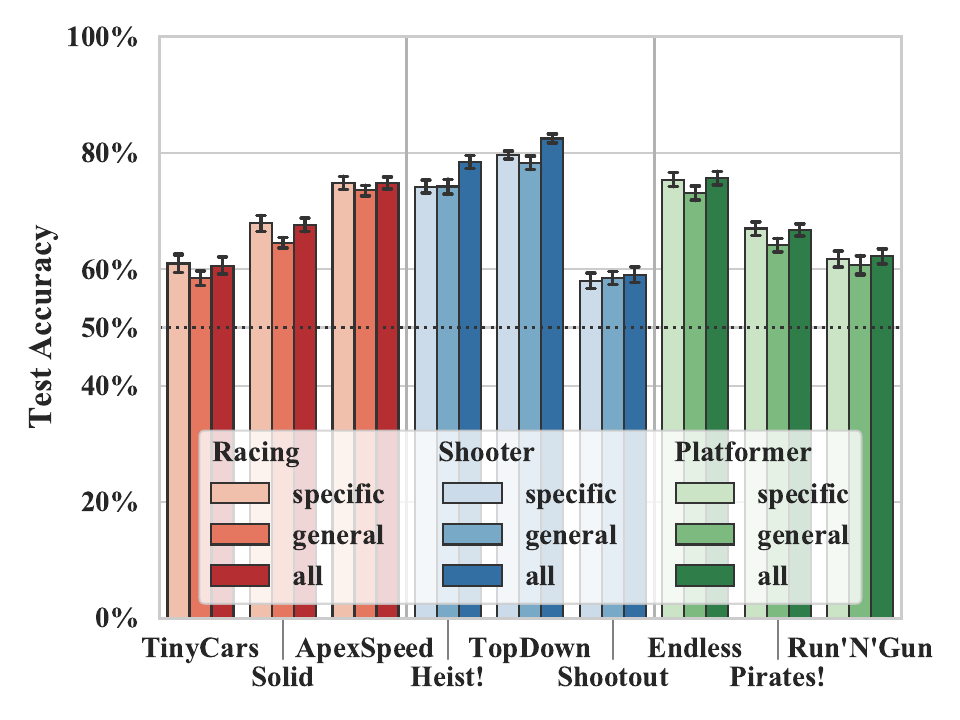}
    \caption{Performance of random forest models of arousal for each game with game-specific, general, and all available features. The dotted line depict the performance baseline and the error bars represent $95\%$ confidence intervals.}
    \label{fig:baselines}
\end{figure}

To examine the validity of the \emph{general features} discussed in Section \ref{sec:features}, models are constructed for each game based on three different feature sets: 1) game-\emph{specific} features excluding general features 2) \emph{general} features across games shown on Table \ref{tab:general_features} and 3) \emph{all} features combined. Due to the pairwise transformation discussed above, the baseline accuracy of all experiments is 50\%. Because RFs are stochastic algorithms, we run each experiment 5 times and we report the 10-fold cross validation accuracy. Note that each fold contains the data of 10 to 12 participants and no two folds contain data from the same participant. The reported statistical significance is measured with two-tailed Student's $t$-tests with $\alpha=0.05$, adjusted with the Bonferroni correction where applicable.

Figure \ref{fig:baselines} shows the performance of the RF models. Prediction accuracy varies between $58.06\%$ and $82.50\%$ across games. The results reveal that arousal appears to be easier to predict in some games (e.g. \emph{ApexSpeed, TopDown}, and \emph{Endless}) than others (e.g. \emph{TinyCars, Shootout}, and \emph{Run'N'Gun}). 
In the racing and platformer genres, games with fewer input options and an automatic progression system (\emph{ApexSpeed} and \emph{Endless} respectively) are tied to higher model performance. An explanation could be that games with more internal structure (due to the sparsity of actions the player can take and automatic progression through the game with minimal input) present a simpler problem. An exception to this observation is \emph{Shootout}, in which the controls are limited (only looking around and shooting) and enemies appearing in an ever-increasing speed, but despite these similarities with \emph{ApexSpeed} and \emph{Endless}, \emph{Shootout} models are struggling to reach 60\% accuracy (the lowest performance across all games).

Looking at individual games across different feature sets, we observe that the \emph{general features} manage to perform comparably to the \emph{specific features} independently of the game tested. Game-specific features yield significantly higher performances than general features only in 4 games (\emph{TinyCars, Solid, Endless}, and \emph{Pirates!}). Moreover, the combination of both specific and general features yields significantly more accurate arousal models than either the game-specific or general features (or both) in 5 games: \emph{Solid, Heist!, TopDown, Endless}, and \emph{Pirates!}. These results demonstrate the robustness of the general features presented in Section \ref{sec:features} and show that there is little to no trade-off in representing the presented games in a more abstract and general manner.

The arousal model performances presented in this section highlight a number of challenges for future research. Firstly, the differences in performances between games show that the complexity of the affect modelling task is dependent on the characteristics of the elicitor and the game context. Finding new processing methods, data treatment, algorithms, and model architectures which perform equally well across different games is an open problem. Secondly, the robustness demonstrated by the general features proposed in this paper point towards the possibility of general player affect modelling across games. While research has already been investigating general player modelling in video games \cite{camilleri2017towards}, early results showed only moderate success. The dataset and baselines presented in this paper provide a large open source database of games with robust enough general features to continue the exploration of general player modelling.

\section{Discussion}\label{sec:discussion}

This paper presented the AGAIN dataset, a database for affect modelling in video games. The dataset contains data from $124$ players and includes game telemetry, gameplay videos, and \emph{arousal} annotations of $1,116$ gameplay sessions. The paper also presented the dataset, discussed the underlying trends in the data, and showcased some preliminary preference learning models. In this section, we discuss some of the limitations and propose avenues for future work, before concluding the paper.

\subsection{Limitations}\label{sec:limitation}

While the crowd-sourcing protocol for data collection enabled a larger dataset with high extensibility potential, most of the limitations of AGAIN stem from the same crowd-sourced protocol. Collected data lacks modalities traditionally associated with AC datasets. Neither physiological signals nor facial expression data is collected, as the online procedure focused on behavioural telemetry instead. While AGAIN features no peripheral signals, the dataset also contains over $37$ hours of gameplay video footage, which can support a number of computer vision-based applications \cite{makantasis2019pixels,makantasis2021pixels}.

Whereas many affective datasets are composed of multiple affective labels---with \emph{arousal} and \emph{valence} being the most common---the AGAIN dataset focuses only on \emph{arousal}. As the game-playing task is already a lengthy and involved process, annotating multiple affective dimensions was infeasible during data collection. The choice of \emph{arousal} was motivated by this affective dimension's strong connection to the dynamics of gameplay. This is especially important for first-person annotations, as games already encode positive and negative events (in the form of helpful and detrimental effects to the goal of the game) which can be assessed by third-person annotators later. However, the subjectively perceived dynamics of the game might differ from an observer's impressions.

While each game elicits similar playstyles across different participants, the database features unique videos with self-annotated arousal traces. AGAIN puts an emphasis on self-reported labels as it is expected to yield ground truths of affect that are closer to the experience \cite{metallinou2013annotation,afzal2014emotion,yannakakis2018ordinal}. The existing in-game footage of AGAIN, however, can be used directly for third-person annotation in future studies. Regardless of the annotation scheme used (first vs. third person) AGAIN annotations are captured in an unbounded fashion which eliminates high degrees of reporting bias \cite{yannakakis2018ordinal,yannakakis2017ordinal}. 

A necessary but limiting factor is that collection of affect labels is not concurrent with the collection of video and telemetry data. Collecting reliable first-person affect labels simultaneously to video game play is impossible due to the high cognitive demand of the task; however, the stimulated recall technique applied here does pose some limitations to the annotation process as certain temporal biases can arise.

Finally, many game sessions in the dataset have an overall increasing intensity, which is reflected in the corresponding affective labels as well. While this is a limitation of the dataset, it is also a limitation of the domain. Although the same increasing intensity is true to even high-budget console and PC games on the macro level, this dynamic is especially true to short casual and mobile games played over short periods.

We note that the presented machine learning models are quite preliminary, aiming to showcase a use-case for the dataset along with a proposed cleaning and modelling pipeline. Future studies should look into training and tuning more complex models on AGAIN. However, as the published dataset contains both the clean and raw data, future work can propose different processing methods. While the presented models show some level of robustness, they do not use all of the data the dataset has to offer, such as the captured videos.

\subsection{Future Work}

The AGAIN dataset was created to facilitate games user modelling through the lens of affective computing. The dataset allows for the adoption of machine learning techniques that use game telemetry and video data to model player arousal. While a more traditional approach was presented here, future studies should utilise the available video database and apply deep leaning methods to create more complex models. As the dataset contains a large set of games, AGAIN is especially useful for research into general affect models. Future studies should focus on the transferability of models across different games and genres in the dataset \cite{melhart2021towards}.
The current dataset only encodes one affective dimension, \emph{arousal}, across videos from nine games; AGAIN, however is easily scalable to more affective dimensions and more game-based affect stimuli. Future work will focus on expanding the labels with expert annotations of valence and dominance to match the format of other affective computing databases \cite{koelstra2012deap,ringeval2013reloca,zafeiriou2017affwild,kossaifi2019sewa}. Its accessibility and its unobtrusive data collection via crowdsourcing make AGAIN easily extendable to more affect labels, affect elicitors and participants. 

\section{Conclusion}\label{sec:conclusion}
This paper introduced a new database for affect modelling, the AGAIN dataset. AGAIN is the largest and most diverse publicly available dataset coupling gameplay context, gameplay videos, and annotated affect to date. It includes a variety of interactive elicitors, in the form of nine games from three popular yet dissimilar game genres. In particular, the dataset consists of 37 hours of video footage accompanied by telemetry and self-annotated \emph{arousal} labels from $1,116$ gameplay sessions played by 124 participants. The motivation behind the construction of this dataset is to facilitate and further advance research on general player modelling through a clean, large-scale, diverse (elicitor-wise) and accessible database. 

Inspired by recent work on the importance of gameplay context as a predictor of affect \cite{makantasis2019pixels}, the user modalities of AGAIN are currently limited to in-game video footage and behavioural telemetry data. In addition, the protocol of AGAIN limits the user modalities available so that crowdsourcing of self-reported affect annotations is both feasible and efficient. While AGAIN puts an emphasis on accessibility---soliciting game context and behavioural data from users as its modalities---the AGAIN games can be used for small-scale, lab-based affect studies that incorporate more user modalities including visual and auditory player cues (e.g. \cite{karpouzis2015platformer,makantasis2021pixels}).

Given the characteristics of a unique set of diverse elicitors, a large participant count, first-person annotations and a large-scale video and game telemetry database, AGAIN couples important aspects of affective computing with core aspects of game user modelling---thereby enabling research in the area of general player modelling, in games and beyond.

\section*{Acknowledgements}

This project has received funding from the European Union’s Horizon 2020 programme under grant agreement No 952002.

\bibliography{again_dataset.bib}
\bibliographystyle{IEEEtran}

\begin{IEEEbiography}[{\includegraphics[width=0.9in,clip,trim={75px 0 75px 0},keepaspectratio]{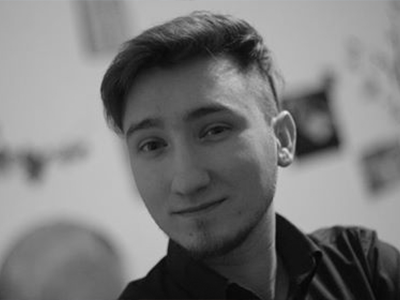}}]
{David Melhart} is a Postdoctoral Researcher at the Institute of Digital Games, University of Malta. He received a MA degree in Cognition and Communication from the University of Copenhagen in 2016 and a Ph.D. degree in Game Research from the University of Malta in 2021. His research focuses on Machine Learning, Affective Computing, and Games User Modelling. He was the Communication Chair of FDG 2020, has been a recurring organiser and Publicity Chair of the \emph{Summer School series on Artificial Intelligence and Games} (2018-2022), and currently serves as an Editorial Assistant to the \emph{IEEE Transactions on Games}.
\end{IEEEbiography}

\begin{IEEEbiography}
[{\includegraphics[width=0.9in,clip,trim={250px 0 50px 0},keepaspectratio]{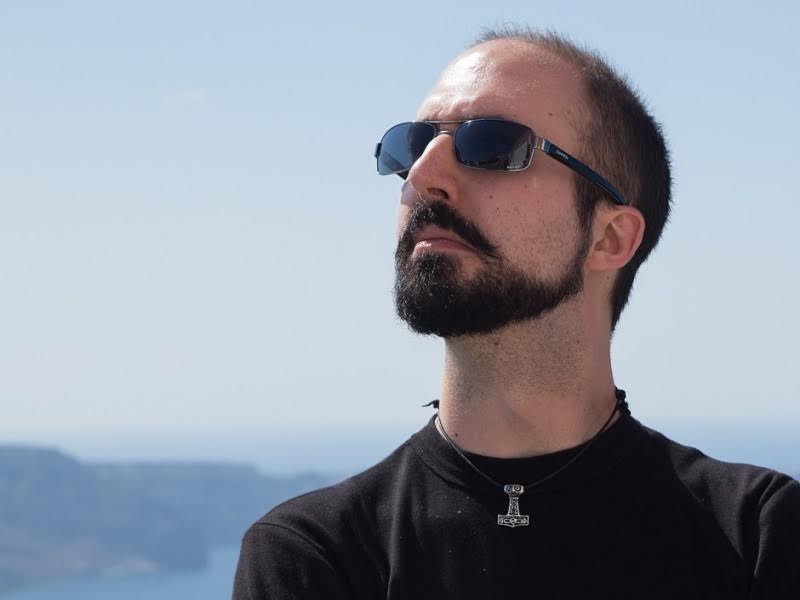}}]
{Antonios Liapis} is a Senior Lecturer at the Institute of Digital Games, University of Malta, where he bridges the gap between game technology and game design in courses focusing on human-computer creativity, digital prototyping and game development. He received the Ph.D. degree in Information Technology from the IT University of Copenhagen in 2014. His research focuses on Artificial Intelligence in Games, Human-Computer Interaction, Computational Creativity, and User Modelling. He has published over 120 papers in the aforementioned fields, and has received several awards for his research contributions and reviewing effort. He serves as Associate Editor for the {IEEE Transactions on Games}, and has served as general chair in four international conferences, as guest editor in four special issues in international journals, and has co-organised 12 workshops. 
\end{IEEEbiography}

\begin{IEEEbiography}[{\includegraphics[width=0.9in,clip,trim={50px 0 100px 0},keepaspectratio]{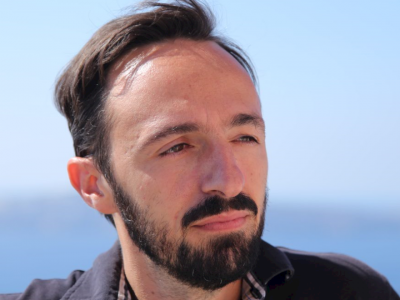}}]
{Georgios N. Yannakakis} is a Professor and Director of the Institute of Digital Games, University of Malta (UM). He received the Ph.D. degree in Informatics from the University of Edinburgh in 2006. Prior to joining the Institute of Digital Games, UM, in 2012 he was an Associate Professor at the Center for Computer Games Research at the IT University of Copenhagen. He does research at the crossroads of artificial intelligence, computational creativity, affective computing, advanced game technology, and human-computer interaction. He has published more than 300 papers in the aforementioned fields and his work has been cited broadly. His research has been supported by numerous national and European grants (including a Marie Skłodowska-Curie Fellowship) and has appeared in \emph{Science Magazine} and \emph{New Scientist} among other venues. He is currently the Editor-in-Chief of the {IEEE Transactions on Games} and an Associate Editor of the {IEEE Transactions on Evolutionary Computation}, and used to be Associate Editor of the {IEEE Transactions on Affective Computing} and the {IEEE Transactions on Computational Intelligence and AI in Games} journals. He has been the General Chair of key conferences in the area of game artificial intelligence (IEEE CIG 2010) and games research (FDG 2013, 2020). Among the several rewards he has received for journal and conference publications he is the recipient of the \emph{IEEE Transactions on Affective Computing Most Influential Paper Award} and the \emph{IEEE Transactions on Games Outstanding Paper Award}. He is a senior member of the IEEE.
\end{IEEEbiography}

\end{document}